\documentclass{paper}

\newcommand{\dd}{{\rm d}}
\newcommand{\Om}{\Omega_{\rm m}}
\newcommand{\Omn}{{\Omega^0_{\rm m}}}
\newcommand{\Ol}{\Omega_\Lambda}
\newcommand{\Oln}{{\Omega^0_\Lambda}}
\newcommand{\V}[1]{\vec{#1}}
\newcommand{\Wb}[1]{ W(\V{r}-\V{r}_{#1},h) }
\newcommand{\Wab}{ W(\V{r}_a-\V{r}_b,h) }
\newcommand{\Wabh}[1]{ W(\V{r}_a-\V{r}_b,#1)}

\begin{document}

\title{SPH simulations of magnetic fields in galaxy clusters}
\author{Klaus Dolag$^1$, Matthias Bartelmann$^1$, and Harald Lesch
  $^2$\\
  $^1$Max-Planck-Institut f\"ur Astrophysik, P.O.~Box 1523,
  D--85740 Garching, Germany\\
  $^2$Universit\"ats-Sternwarte M\"unchen, Scheinerstr.~1, D-81679
  M\"unchen, Germany}

\date{\em accepted by Astronomy \& Astrophysics}

\begin{abstract}
We perform cosmological, hydrodynamic simulations of magnetic fields
in galaxy clusters. The computational code combines the
special-purpose hardware Grape for calculating gravitational
interaction, and smooth-particle hydrodynamics for the gas
component. We employ the usual MHD equations for the evolution of the
magnetic field in an ideally conducting plasma. As a first
application, we focus on the question what kind of initial magnetic
fields yield final field configurations within clusters which are
compatible with Faraday-rotation measurements. Our main results can be
summarised as follows: (i) Initial magnetic field strengths are
amplified by approximately three orders of magnitude in cluster cores,
one order of magnitude above the expectation from spherical
collapse. (ii) Vastly different initial field configurations
(homogeneous or chaotic) yield results that cannot significantly be
distinguished. (iii) Micro-Gauss fields and Faraday-rotation
observations are well reproduced in our simulations starting from
initial magnetic fields of $\sim10^{-9}\,{\rm G}$ strength at redshift
15. Our results show that (i) shear flows in clusters are crucial for
amplifying magnetic fields beyond simple compression, (ii) final field
configurations in clusters are dominated by the cluster collapse
rather than by the initial configuration, and (iii) initial magnetic
fields of order $10^{-9}\,{\rm G}$ are required to match
Faraday-rotation observations in real clusters.

\end{abstract}

\maketitle

\section{\label{sec:1} Introduction}

Magnetic fields in galaxy clusters are inferred from observations of
diffuse radio haloes (Kronberg 1994), Faraday rotation (Vallee,
MacLeod \& Broten 1986, 1987), and recently also hard X-ray emission
(Bagchi, Pislar \& Lima Neto 1998). The diffuse radio emission comes
from the entire clusters rather than from individual radio sources. It
is typically unpolarised and has a power-law spectrum, indicative of
synchrotron radiation from relativistic electrons with a power-law
energy spectrum in a magnetic field. Since the magnetised
intra-cluster plasma is birefringent, it gives rise to Faraday
rotation, which is detectable through multi-frequency radio
observations of polarised radio sources in or behind the
clusters. Hard X-ray emission is due to CMB photons which are
Compton-upscattered by the same relativistic electron population
responsible for the synchrotron emission. {\em Upper\/} limits to the
hard X-ray emission have previously been used to infer {\em lower\/}
limits to the magnetic field strengths: Stronger fields require less
electrons to produce the observed radio emission, and this electron
population is less efficient in scattering CMB photons to X-ray
energies. Even non-detections of hard X-ray emission (e.g.~Rephaeli \&
Gruber 1988) have therefore been useful to infer that cluster-scale
magnetic fields should at least be of order $1\,\mu{\rm G}$. Combined
observations of hard X-ray emission and synchrotron haloes are most
valuable. They allow to infer magnetic field strengths without further
restrictive assumptions because they are based on the same
relativistic electron population.

It is therefore evident that clusters of galaxies are pervaded by
magnetic fields of $\sim\mu{\rm G}$ strength. Small-scale structure in
the fields has been seen in high-resolution observations of extended
radio sources (Dreher, Carilli \& Perley 1987; Perley 1990; Taylor \&
Perley 1993; Feretti et al.~1995), and inferred from Faraday rotation
measurements in conjunction with X-ray observations and
inverse-Compton limits (Kim et al.~1990). Coherence of the observed
Faraday rotation across large radio sources (Taylor \& Perley 1993)
demonstrates that there is at least a field component that is smooth
on cluster scales. Very small-scale structure and steep gradients in
the Faraday rotation measure (Dreher et al.~1987; Taylor, Barton \& Ge
1994) may partially be due to Faraday rotation intrinsic to the
sources or in cooling flows in cluster centres.

Various models have been proposed for the origin of cluster magnetic
fields in individual galaxies (Rephaeli 1988; Ruzmaikin, Sokolov \&
Shukurov 1989). The basic argument behind such models is that the
metal abundances in the intra-cluster plasma are comparable to solar
values. The plasma therefore must have been substantially enriched by
galactic winds which could at the same time have blown the galactic
magnetic fields into intra-cluster space.

Magnification of galactic seed fields in turbulent dynamos driven by
the motion of galaxies through the intra-cluster plasma appeared for
some time as the most viable model (Ruzmaikin et al.~1989; Goldman \&
Rephaeli 1991). It has been shown recently (De Young 1992; Goldshmidt
\& Rephaeli 1993) that it is difficult for that process to create
cluster-scale fields of the appropriate strengths, mainly because of
the small turbulent velocities driven by galactic wakes, and because
turbulent energy cascades to the dissipation scale more rapidly than
the dynamo process could amplify the field. Field strengths of
$0.1\,\mu{\rm G}$ seem to be the maximum achievable under idealised
conditions, and typical correlation scales are of order the turbulent
scale, $\sim10\,{\rm kpc}$, and below.

It is clear that this dynamo mechanism occurs in galaxy clusters and
produces small-scale structure in the cluster magnetic fields. Other
processes have to be invoked as well in order to explain micro-Gauss
fields on cluster scales.

We here address the question whether primordial magnetic fields of
speculative origin, magnified in the collapse of cosmic material into
galaxy clusters, can reproduce a number of observations, particularly
Faraday-rotation measurements. Specifically, we ask what initial
conditions we must require for the primordial fields in order to
reproduce the statistics of rotation-measure observations. A
compilation of available observations of this kind was published by
Kim, Kronberg \& Tribble (1991).

\begin{figure}[ht]
  \center{\includegraphics[width=\hsize]{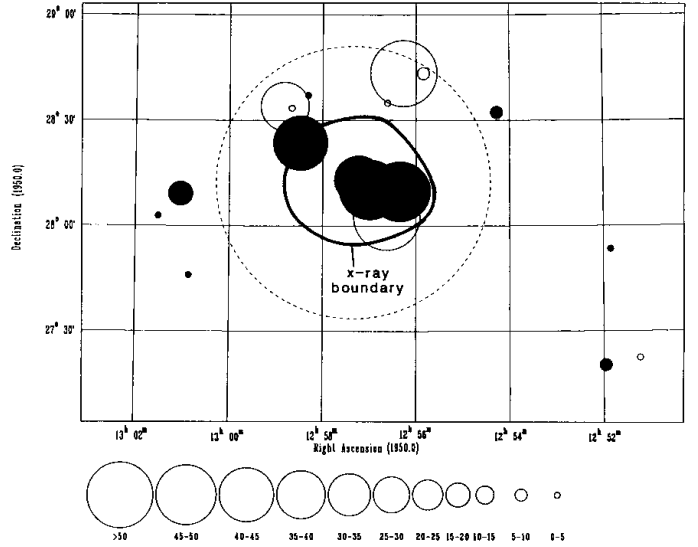}}
\caption{Rotation measures observed in the Coma cluster (reproduced
  from Kim et al.~1990). The size of the circles indicates the
  absolute value of RM, the type (filled or empty) the sign of RM. The
  X-ray boundary (solid curve) and the Abell radius (dashed circle)
  are superposed to mark the position of the Coma cluster in the
  plot. The RM pattern can later be compared with our simulated
  results as shown in Fig.~\ref{fig:3dbox}.}
\label{fig:coma}
\end{figure}

Measurements of Faraday rotation are made possible by the fact that
synchrotron emission in an ordered magnetic field produces linearly
polarised radiation, and therefore the emission of many observed radio
sources is linearly polarised to some degree. The plane of
polarisation is rotated when the radiation propagates through a
magnetised plasma like the intra-cluster medium (ICM). Observing a
background radio source through a galaxy cluster at different
wavelengths, and measuring the position angle $\phi$ of the
polarisation allows to determine the rotation measure,
\begin{eqnarray}
  {\rm RM} &=& \Delta\phi\,
  \frac{\lambda_1^2\lambda_2^2}{\lambda_2^2-\lambda_1^2} =
  \frac{e^3}{2\pi m_{\rm e}^2\,c^4}\int n_{\rm e} B_\parallel
  \dd l\nonumber\\
  &=& 812\,\frac{{\rm rad}}{{\rm m}^2}\,
  \int \frac{n_{\rm e}}{{\rm cm}^{-3}}\,
  \frac{B_\parallel}{\mu{\rm G}}\,\frac{\dd l}{{\rm kpc}}\;.
\end{eqnarray}
It can have either sign, depending on the orientation of the magnetic
field. Figure~\ref{fig:coma} shows all published rotation measures
observed in radio sources seen through the Coma cluster (Kim et
al.~1990). The observed Faraday rotation is averaged across the source
or the telescope beam, whichever is smaller. Most of the sources used
for Faraday-rotation measurements are QSOs or faint radio galaxies,
whose size is negligible compared to the resolution of the simulations
presented below. The number of RM values available per cluster depends
on the number-density of background radio sources bright enough to
measure polarisation. The noise level for reliable Faraday-rotation
measurements is about $25-35\,\mu{\rm Jy}$ for several minutes of
integration time at the VLA. The polarised flux should be about
$300\,\mu{\rm Jy}$, i.e.~the total flux should be about $10~mJy$ for a
source polarised at the level of a few per cent. With a large amount
of observation time one could reach about $20-40$ rotation measures
per square degree (P.P.~Kronberg, private communication). We note that
Kim et al.~(1990) had on average 4 background rotation measures per
square degree for the Coma cluster.

Thanks to recent compilations (Kim et al.~1991; see also Goldshmidt \&
Rephaeli 1993) of large rotation-measure samples, the observational
basis for our exercise now seems to be sufficiently firm.

The numerical method used rests upon the GrapeSPH code written and
kindly provided by Matthias Steinmetz (Steinmetz 1996). The code
combines the merely gravitational interaction of a dark-matter
component with the hydrodynamics of a gaseous component. We supplement
this code with the magneto-hydrodynamic equations to trace the
evolution of the magnetic fields which are frozen into the motion of
the gas because of its ideal electric conductivity. The latter
assumptions precludes dynamo action in our simulations.

\section{\label{sec:2} GrapeSPH combined with MHD}

We start with the GrapeSPH code developed and kindly provided by
Matthias Steinmetz (Steinmetz 1996). The code is specialised to the
``Grape'' (Gravity Pipe) hardware briefly described below. It
simultaneously computes with a multiple time-step scheme the behaviour
of two matter components, a dissipation-free dark matter component
interacting only through gravity, and a dissipational, gaseous
component. The gravitational interaction is evaluated on the Grape
board, while the hydrodynamics is calculated by the CPU of the host
work-station in the smooth-particle approach (SPH). We supplement the
code with the magneto-hydrodynamic equations to follow the evolution
of an initial magnetic field caused by the flow of the gaseous matter
component.

We briefly describe here the Grape board and SPH as far as necessary
for our purposes.

\subsection{GRAPE}

Gravitational forces are calculated on the special-purpose hardware
component Grape 3Af (Ito et al.~1993) which is connected to a
Sun-Sparc~10 work-station operating at 50~MHz. Given a collection of
particles, their masses and positions, the Grape board computes their
mutual distances and the gravitational forces between them, smoothed
at small distances according to the Plummer law. A fixed particle
(index $a$) at position $\V{r}_a$ with velocity $\V{v}_a$ thus
experiences the gravitational acceleration
\begin{eqnarray}
  \left(\frac{\dd\V{v}_a}{\dd t}\right)_\mathrm{grav} &=&
  \sum_i\frac{m_i}
  {(|\V{r}_a-\V{r}_i|^2+0.25(\epsilon_a+\epsilon_i)^2)^{1.5}}
  \nonumber\\
  &\times&(\V{r}_a-\V{r}_i)\;,
\label{eq:1}
\end{eqnarray}
with the sum running over all other particles $i$ at positions
$\V{r}_i$ and with masses $m_i$. The softening length $\epsilon$ is
adapted to the mass of each particle to prevent floating-point
overflows beyond the restricted dynamical range of the Grape hardware.
Conveniently, the Grape board also returns a list of neighbouring
particles to a given particle, which is helpful later on for the SPH
part of the code.

\subsection{SPH}

Smoothed-particle hydrodynamics (SPH; Lucy 1977; Monaghan 1985)
replaces test particles by spheres with variable radii. An ensemble of
such extended particles with masses $m_i$ at positions $\V{r}_i$ then
gives rise to the average mass density at position $\V{r}$,
\begin{equation}
  \langle\rho(\V{r})\rangle = \sum_i\,m_i\,W(\V{r}-\V{r}_i,h)\;.
\label{eq:2}
\end{equation}
The spatial extent of the particles is specified by the SPH kernel
$W(\delta\V{r},h)$ which depends on the distance $\delta\V{r}$ from
the point under consideration and has a finite width parameterised by
$h$. The kernel must be normalised, and approach a Dirac delta
distribution for vanishing $h$. A normalised Gaussian fulfils these
conditions, but for numerical purposes kernels with compact support
are preferred. We use the $B_2$-spline kernel (Monaghan 1985). The
summation over all particles is then effectively reduced to a
summation over neighbouring particles.

Physical quantities $A$ are calculated at the particle positions
$\V{r}_b$ and extrapolated to any desired position $\V{r}$ in an
analogous manner,
\begin{equation}
  \langle A(\V{r})\rangle = \sum_b\,m_b\,
  \frac{A(\V{r}_b)}{\rho(\V{r}_b)}\,W(\V{r}-\V{r}_b,h)\;,
\label{eq:3}
\end{equation}
where $b$ runs over all particles within a neighbourhood of $\V{r}$
specified by $h$, and $m_b\,\rho^{-1}(\V{r}_b)$ represents the volume
element occupied by the $b\,$th particle. The crucial step in SPH is
to identify the averaged quantity $\langle A(\V{r})\rangle$ from
eq.~(\ref{eq:3}) with the physical quantity $A(\V{r})$.

According to eq.~(\ref{eq:3}), spatial derivatives of $A(\V{r})$ can
be expressed as sums over $A(\V{r}_b)$ times spatial derivatives of
the kernel function, which are analytically known from the
start. Thus,
\begin{equation}
  \nabla A(\V{r}) = \sum_b\,m_b\,
  \frac{A(\V{r}_b)}{\rho(\vec{r}_b)}\,\nabla\Wb{b}\;.
\label{eq:4}
\end{equation}
The hydrodynamic equations are correspondingly simplified. Here we
need the derivatives at a particle position $\V{r}_a$. We therefore
replace $\V{r}$ by $\V{r}_a$ and mark the gradient operator with a
subscript $a$ to indicate that the gradient is to be taken with
respect to $\V{r}_a$. The momentum equation can then be written
\begin{eqnarray}
  \left(\frac{\dd\V{v}_a}{\dd t}\right)_\mathrm{hyd} &=&
    \sum_b\,m_b\,\left(
    \frac{P_b}{\rho_b^2}+\frac{P_a}{\rho_a^2}+\Pi_{ab}
  \right)\nonumber\\
  &\times&\nabla_a\Wab\;,
\label{eq:5}
\end{eqnarray}
and the equation for the internal energy $u_a$ takes the form
\begin{eqnarray}
  \frac{\dd u_a}{\dd t} &=& \sum_b\,m_b\,\left(
    \frac{P_b}{\rho_b^2}+\frac{1}{2}\,\Pi_{ab}
  \right)\,(\V{v}_a-\V{v}_b)\nonumber\\
  &\times&\nabla_a\Wab\;.
\label{eq:6}
\end{eqnarray}
These equations need to be supplemented by an equation of state. We
assume an ideal gas with an adiabatic index of $\gamma=5/3$, for which
the pressure is given by $P_i=(\gamma-1)\,u_i\rho_i$. The tensor
$\Pi_{ij}$ describes an artificial viscosity term required to properly
capture shocks. We adopt the form proposed by Monaghan \& Gingold
(1983) and Monaghan (1989) which includes a bulk-viscosity and a
von~Neumann-Richtmeyer viscosity term, supplemented by a term
controlling angular-momentum transport in presence of shear flows at
low particle numbers (Balsara 1995, Steinmetz 1996).

The code automatically adapts the spatial kernel
$\Wab=0.5[\Wabh{h_a}+\Wabh{h_b}]$ and its width $h_i$ for each
particle in such a way that the number of neighbouring particles falls
between 50 and 80. This results in an adaptive spatial resolution
which depends on the mass of the SPH particle and the local mass
density at the particle position, because the number of neighbours is
determined by the local density and the particle mass.

\subsection{MHD}

For an ideally conducting plasma, the induction equation can be
written as
\begin{equation}
  \frac{\dd\V{B}}{\dd t} = (\V{B}\cdot\nabla)\V{v} -
  \V{B}(\nabla\cdot\V{v})+\V{v}(\nabla\cdot\V{B})\;.
\label{db}
\end{equation}
Theoretically of course, the last term in eq.~(\ref{db}) can be
ignored. Numerically however, $\nabla\cdot\V{B}$ will not exactly
vanish, and therefore the question arises whether the {\em actual\/}
or the {\em ideal\/} divergence should be inserted when numerically
evaluating the induction equation. We performed tests showing that
including $\nabla\cdot\V{B}$ in eq.~(\ref{db}) results in a strong
increase of any initial $\nabla\cdot\V{B}$ as time proceeds. When we
set $\nabla\cdot\V{B}=0$, however, an initially small but
non-vanishing divergence rises at most in proportion to $|\V{B}|$,
thus remaining negligibly small.

The field acts back on the plasma with the Lorentz force
\begin{equation}
  \V{F} = -\nabla\left(\frac{\V{B}^2}{8\pi}\right) +
  \frac{1}{4\pi}(\V{B}\cdot\nabla)\V{B}\;.
\label{fb}
\end{equation}
Formally inserting $\nabla\cdot\V{B}$ into this expression, the
magnetic force can be written as the divergence of a tensor
${\cal M}_{ij}$,
\begin{equation}
  F_j = \frac{\partial{\cal M}_{ij}}{\partial x_i}\;,
\label{eq:9}
\end{equation}
with components of ${\cal M}$ given by
\begin{equation}
  {\cal M}_{ij} = \frac{1}{4\pi}\,\left(
    \V{B}_i\V{B}_j-\frac{1}{2}\V{B}^2\delta_{ij}
  \right)\;.
\label{mwt}
\end{equation}
The two formally equivalent expressions (\ref{fb}) and (\ref{eq:9})
for the magnetic force have different advantages.

Manifestly in conservation form, eq.~(\ref{eq:9}) conserves linear and
angular momenta exactly. However, for strong magnetic fields,
i.e.~fields for which the Alfv{\'e}n speed is comparable or larger
than the sound speed, motion can become unstable (Phillips \& Monaghan
1985). In our cosmological cluster simulations, magnetic fields never
reach such high values, so we can safely employ eq.~(\ref{eq:9}). We
compared cluster simulations performed with both force equations
(\ref{fb}) and (\ref{eq:9}), finding no significant differences in the
resulting magnetic field. Nonetheless, we switched to eq.~(\ref{fb})
for the tests described below in which strong magnetic fields
occur. As expected, results are improved in such cases.  This is
partially due to the fact that the shock tube problem described in
Sect.~\ref{testing} has an unphysical divergence in $\V{B}$ at the
shock.

In the language of SPH, eqs.~(\ref{db}) and (\ref{eq:9}) read
\begin{eqnarray}
  \frac{\dd \V{B}_{a,j}}{\dd t} &=& \frac{1}{\rho_a}\,
  \sum_b\,m_b\,(\V{B}_{a,j}\V{v}_{ab}-\V{B}_a\V{v}_{ab,j})
  \nonumber\\&\times&
  \nabla_a\Wab\;,
\label{eq:10}
\end{eqnarray}
and
\begin{eqnarray}
  \left(\frac{\dd\V{v}_a}{\dd t}\right)_\mathrm{mag} &=&
  \sum_b\,m_b\,\left[
    \left(\frac{{\cal M}}{\rho^2}\right)_a+
    \left(\frac{{\cal M}}{\rho^2}\right)_b
  \right]\,\nonumber\\
  &\times&\nabla_{a}\Wab\;.
\label{eq:11}
\end{eqnarray}
Here, comma-separated indices $j$ mean the $j\,$th component of any
vector. Note that $\nabla_a\Wab$ is a vector and therefore leads to a
scalar product in eq.~(\ref{eq:10}), and to a matrix product with
${\cal M}$ in eq.~(\ref{eq:11}).

\section{Initial conditions}

We need two types of initial conditions for our simulations, namely
(i) the cosmological parameters and initial density perturbations, and
(ii) the properties of the primordial magnetic field. We detail our
choices here.

\subsection{Cosmology}

For the purposes of the present study, we set up cosmological initial
conditions in an Einstein-de Sitter universe ($\Omn=1$, $\Oln=0$) with
a Hubble constant of $H_0=50\,{\rm km\,s^{-1}\,Mpc^{-1}}$. We
initialise density fluctuations according to a COBE-normalised CDM
power spectrum.

The mean matter density in the initial data is determined by the
current density parameter $\Omn$. The Hubble expansion, calibrated by
the Hubble constant $H_0$, is represented by an isotropic velocity
field, which is appropriately added to the initial peculiar velocities
of the simulation particles. We use $H_0$ in units of $100\,{\rm
km\,s^{-1}\,Mpc^{-1}}$ as dimensionless Hubble constant instead of the
conventional $h$ to avoid confusion with the width of the SPH
kernel. A cosmological constant $\Oln$ could be introduced by adding
the term
\begin{equation}
  \left(\frac{\dd\V{v}_a}{\dd t}\right)_\mathrm{cos} =
  \Oln\,H_0^2\,\V{x}_a
\label{eq:12}
\end{equation}
to the force equation, according to Friedmann's equations.

Density fluctuations normalised such as to reproduce the local
abundance of galaxy clusters (White, Efstathiou \& Frenk 1993) can be
mimicked by interpreting simulations at redshift $z>0$ as
corresponding to $z=0$.  We then have to use $H(z)$ instead of $H_0$,
and the normalisation is reduced by $(1+z)^{-1}$. $\Om(z)$ and
$\Ol(z)$ do not change with redshift in an Einstein-de Sitter
universe.

\subsection{\label{primfields} Primordial magnetic fields}

The origin of the observed cluster magnetic fields is still
unclear. As mentioned in the introduction, various models have been
proposed which relate the cluster fields with the field generation in
the individual galaxies and subsequent wind-like activity which
transports and redistributes the magnetic fields in the intra-cluster
medium (e.g.~Ruzmaikin, Sokolov \& Shukurov 1989; Kronberg, Lesch \&
Hopp 1999). Such scenarios are connected with the observations of
metal-enriched material in the cluster, which presume a significant
enrichment by galactic winds. However, Goldshmidt \& Rephaeli (1994)
give strong arguments against intracluster magnetic fields being
expelled from cluster galaxies.

Alternatively, it has been proposed that the intergalactic magnetic
fields may be due to some primordial origin in the pre-recombination
era (Rees 1987) or pre-galactic era (Lesch \& Chiba 1995; Wiechen,
Birk \& Lesch 1998). The first class of models can be disputed on the
grounds of principal plasma-physical objections concerning the
electric conductivity at the very high temperature stages of the very
early universe. The proton-photon and electron-photon collisions
produce such a high resistance that primordial magnetic fields are
likely not to survive (Lesch \& Birk 1998).

The second category (the pre-galactic models) do not reach sufficient
field strengths to account for the micro-Gauss fields observed in the
intra-cluster medium. Thus we are left with the wind mechanism, which
transports magnetic flux into the cluster medium. The initial
configuration depends a lot on the time scale on which the magnetic
flux is transported into forming cluster structures. If galaxies,
especially dwarf galaxies, are formed much earlier than galaxy
clusters, they can generate and redistribute magnetic fields very
early. This would probably lead to large-scale magnetic fields of
about $10^{-9}\,{\rm G}$ on Mpc scales (Kronberg, Lesch \& Hopp
1999). If the cluster fields are due to winds from galaxies which
later become part of the cluster, the initial field configuration will
not be that simple. Taking into account the high velocity dispersion
of individual galaxies in clusters and the high electrical
conductivity of the plasma, we must expect that the initial cluster
field exhibited a very chaotic structure and only the mass flow into
and within the cluster can order and amplify it to the observed field
strengths.

The simulations presented here start with both set-ups of the
primordial magnetic field, namely either a chaotic or a completely
homogeneous magnetic field at high redshift. The average magnetic
field energy is fixed to the same value in both cases to allow a fair
comparison of the results.

The homogeneous magnetic field can be superposed in an arbitrary
spatial direction, for which we have chosen the direction of one of
the coordinate axes.

To set up the random field, we draw Fourier components of the field
strengths, $|\V{B}_k|$, from a power spectrum of the form $P_{\rm
B}(k)=A_0\,k^\alpha$. That is, the $|\V{B}_k|$ are drawn from a
Gaussian distribution with mean zero and standard deviation $P_{\rm
B}^{1/2}(k)$. To completely specify $\V{B}$ under the constraint
$\nabla\cdot\V{B}=0$, we choose a random orientation for the wave
vector $\V{k}$ and components of $\V{B}_k$ such as to satisfy
$\V{k}\cdot\V{B}_k=0$. We set the power-spectrum exponent $\alpha=5/3$
corresponding to a Kolmogorov spectrum (Biskamp 1993).

We also vary the strength of the initial magnetic field. We use the
average initial field energy density $\langle\V{B}^2\rangle_{\rm
ini}/4\pi$ to parameterise initial magnetic field strengths. The
magnetic fields are set up at the initial redshift of the simulations,
$z=15$.

\subsection{Simulation Parameters}

Our simulations work with three classes of particles. In a central
region, we have $\sim50,000$ collisionless dark-matter particles with
mass $3.2\times10^{11}\,M_\odot$, mixed with an equal number of gas
particles whose mass is twenty times smaller. This is the region where
the clusters form. At redshift $z=15$ where the simulations are set
up, it is is a sphere with comoving diameter $\sim4.5\,{\rm Mpc}$. The
central region is surrounded by $\sim20,000$ collisionless boundary
particles whose mass increases outward to mimic the tidal forces of
the neighbouring large scale structure. Including the region filled
with boundary particles, the simulation volume is a sphere with
(comoving) diameter $\sim20\,{\rm Mpc}$ at $z=15$.

As mentioned before, the SPH spatial resolution depends on the local
mass density. For example, the SPH kernel width $h$ is reduced to
$h\approx100\,{\rm kpc}$ in the centres of clusters at redshift
$z=0$. The mean interparticle separation near cluster centres is of
order $10\,{\rm kpc}$.

We use a previously constructed sample of eleven different
realisations of the initial density field (Bartelmann, Steinmetz \&
Weiss 1995), to which we add different initial magnetic-field
configurations as described before. These density fields were taken
out of a large simulation box with COBE-normalised CDM perturbation
spectrum (Bardeen et al.~1986) at places where clusters formed in
later stages of the evolution. They were re-calculated after adding
small-scale power to the initial configurations, taking the tidal
fields of the surrounding matter into account.

Generally, we use only the most massive object in the central region
of the simulation volume, but for some purposes to be described later,
we include up to ten of the next most massive objects found there. The
objects are characterised by their virial radii $R_{200}$, which are
the radii of spheres within which the average mass density is 200
times the critical density $\rho_{\rm crit}$. The virial mass is then
given by $M_{200}=(800\pi/3)\,R_{200}^3\rho_{\rm crit}$. All
quantities in the code are expressed in physical units. When
convenient, we multiply with $H_0$ to convert to the conventional
units with respect to the dimensionless Hubble constant $H_0^{-1}$.

\section{Tests of the code}

In order to test the code, we follow two strategies. The first
consists in re-calculating several test problems, the second in
monitoring some quantities in the process of our cosmological
simulations, e.g.~the $\nabla\cdot\V{B}$ term.

\subsection{\label{testing} Co-planar MHD Riemann problem}

A major test of the code is the solution of a co-planar MHD Riemann
problem. We compare our results with those obtained with a variety of
other methods by Brio \& Wu (1988). This problem is a one-dimensional
shock tube problem with a two-dimensional magnetic field. For that
comparison, we restrict our code to one dimension in space and
velocity, and two dimensions in the magnetic field. In addition, we
switch off gravitational interactions between particles. We choose 400
SPH particles, which is the same number as Brio \& Wu had grid points.

In general, our code clearly and accurately reproduces the results
shown by Brio \& Wu. In particular, we recover all separate stationary
states and waves with correct positions and amplitudes moving into the
right directions. However, we also observe particles oscillating
around the magnetic shock. The reason for that is that a magnetic
discontinuity is set up in such a way that particles, which are driven
by the pressure gradient at the shock, are always accelerated towards
the discontinuity by the magnetic force. It therefore happens that SPH
particles can pass each other or bounce about the discontinuity. This
behaviour adds additional high-level noise on top of some of the
steady states formed on the low-density side of the shock.

This happens only because the magnetic field is strong enough to
produce a magnetic force near the shock which sometimes exceeds the
force created by the pressure gradient. We never encountered a
comparable situation in our cosmological simulations, where the
magnetic force is typically very weak compared to the
pressure-gradient or gravitational forces. Such an oscillatory
behaviour is therefore not expected to happen in cosmological
applications of our code.

\subsection{Divergence of $\V{B}$}

As mentioned earlier, our code does not encounter problems with
$\nabla\cdot\V{B}$ as long as we ignore any non-vanishing
$\nabla\cdot\V{B}$ terms in the induction equation~(\ref{db}). The
initial magnetic field is specified at SPH particle positions. When
extrapolated to grid points, the field acquires a small, but
non-vanishing divergence in $\V{B}$. This initial divergence grows in
proportion with $|\V{B}|$ in the mean, thus remaining negligibly
small. Table~\ref{tab:divb} gives some typical values encountered in
our simulations.

\begin{table}
\caption{The growth of the divergence of the magnetic field is
  illustrated here for one representative cluster simulation with
  initially homogeneous magnetic field. The 25, 50, and 75 percentiles
  of the cumulative distribution of $|\nabla\cdot\V{B}|$ are given for
  three simulation redshifts $z$. During that time, the magnetic field
  near the cluster centre rises from $\sim10^{-9}\,{\rm G}$ to
  $\sim10^{-6}\,{\rm G}$, while the 50-percentile (i.e.~the median of
  $|\nabla\cdot\V{B}|$) increases from $\sim2.1\times10^{-35}\,{\rm
  G\,cm}^{-1}$ to $\sim1.7\times10^{-33}\,{\rm G\,cm}^{-1}$.}
\label{tab:divb}
\medskip
\begin{center}
\begin{tabular}{|r|rrr|}
\hline
  $z$ &
  $|\nabla\cdot\V{B}|_{25\%}$ &
  $|\nabla\cdot\V{B}|_{50\%}$ &
  $|\nabla\cdot\V{B}|_{75\%}$ \\
  & \multicolumn{3}{c|}{in ${\rm G\,cm}^{-1}$} \\
\hline
 1.7 & $1.6\,10^{-36}$ & $2.1\,10^{-35}$ & $2.9\,10^{-34}$ \\
 0.9 & $1.1\,10^{-35}$ & $1.4\,10^{-34}$ & $2.1\,10^{-33}$ \\
 0.0 & $1.2\,10^{-34}$ & $1.7\,10^{-33}$ & $2.3\,10^{-32}$ \\
\hline
\end{tabular}
\end{center}
\end{table}

To relate field strengths and divergences, field strengths need to be
divided by a typical length scale, e.g~the correlation length of the
field. Typically, this is of order $l\sim50-70\,{\rm kpc}$ in our
simulations. Therefore, $B/l\sim5\times10^{-30}\,{\rm G\,cm}^{-1}$,
approximately three orders of magnitude larger than the median
$|\nabla\cdot\V{B}|$. We also checked the fraction of the cluster
volume (defined by a sphere of virial radius) occupied by regions with
the largest 10\% of $\nabla\cdot\V{B}$ and found it to be only
$\sim3\%$.

\section{\label{sec:3} Qualitative results}

\subsection{Importance of shear flows}

The magnetic field in our simulated clusters is dynamically
unimportant even in the densest regions, i.e.~the cluster cores. Since
we ignore cooling, this result may change close to cluster centres
where cooling can become efficient and cooling flows can form.

The amplification of the magnetic field in a volume element during
cluster formation depends on the direction of the magnetic field
relative to the motion of the volume element. If the volume element is
compressed along the magnetic field line, the strength of the magnetic
field is not enhanced at all. If it is compressed perpendicular to the
magnetic field lines, the magnetic field strength is enhanced since
the number of magnetic field lines per unit volume is increased.

The expected growth of the magnetic seed field due to the formation of
a cluster by gravitational collapse can be estimated by the evolution
of a randomly magnetised sphere, spherically collapsing due to
gravity. Flux conservation leads to an enhancement of the magnetic
field proportional to $\rho^{2/3}$. For galaxy clusters, typical
overdensities are of order $10^3$. This means that starting with
magnetic field strengths of order $10^{-9}\,{\rm G}$, spherical
collapse can only produce fields of $10^{-7}\,{\rm G}$ strength, an
order of magnitude below the $\mu{\rm G}$ fields required to explain
observational results.

In our simulations, however, seed fields with $10^{-9}\,{\rm G}$ are
indeed amplified to reach $10^{-6}\,{\rm G}$ in final stages of
cluster evolution. This is due to shear flows which stretch the
magnetic field and amplify it via formation of localised field
structures with enhanced strengths. This local amplification is often
related to Kelvin-Helmholtz instabilities which form magnetic
filaments with much higher field strengths than the surrounding
environment. A necessary condition for such an amplification is that
the relative velocity on both sides of the boundary layer exceeds the
local Alfv\'en speed which is proportional to the field strength, so
that the condition is easy to satisfy for weak initial fields. For
galaxy clusters this effect was first discussed by Livio, Regev \&
Shaviv (1980) and recently simulated in the context of magnetic fields
in galaxy haloes by Birk, Wiechen \& Otto (1998) who could show that
magnetic field strengths are indeed enhanced by a factor of ten by the
Kelvin-Helmholtz instability.

\begin{figure}[ht]
  \center{\includegraphics[width=\hsize]{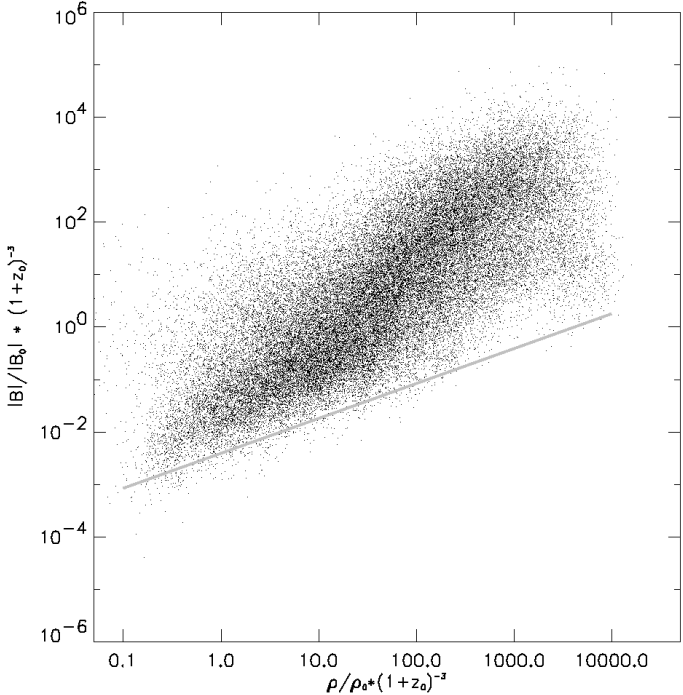}}
\caption{Illustration of the growth of the magnetic field in the
  simulation volume. For each SPH particle at redshift zero, the
  absolute magnetic field $|\V{B}(0)|$ is plotted against the particle
  density $\rho(0)$. Both $|\V{B}(0)|$ and $\rho(0)$ are scaled by
  their values at the initial redshift $z=15$. The line shows the
  expected $(\rho/\rho_0)^{2/3}$ behaviour for the magnetic field
  growth in an isotropic collapse of a randomly magnetised
  sphere. Evidently, the field growth exceeds the simple collapse
  prediction for almost all particles.}
\label{fig:bevol}
\end{figure}

Figure~\ref{fig:bevol} illustrates the growth of the magnetic field in
a representative cluster. It shows the magnetic field at the position
of all SPH particles compared to the mass density at the particle
position.

\subsection{Field Structure}

As mentioned before, overdensities are a necessary condition for the
amplification of magnetic fields, but their presence is not
sufficient: Regions with high density but low magnetic field can also
occur, depending on the relative directions of compression and
magnetic field in a volume element. In addition, the field is also
amplified by shear flows. It is therefore not expected that the
magnetic field in a cluster acquires an orderly radial profile like,
e.g.~the density.

\begin{figure*}[ht]
  \center{\includegraphics[width=\hsize]{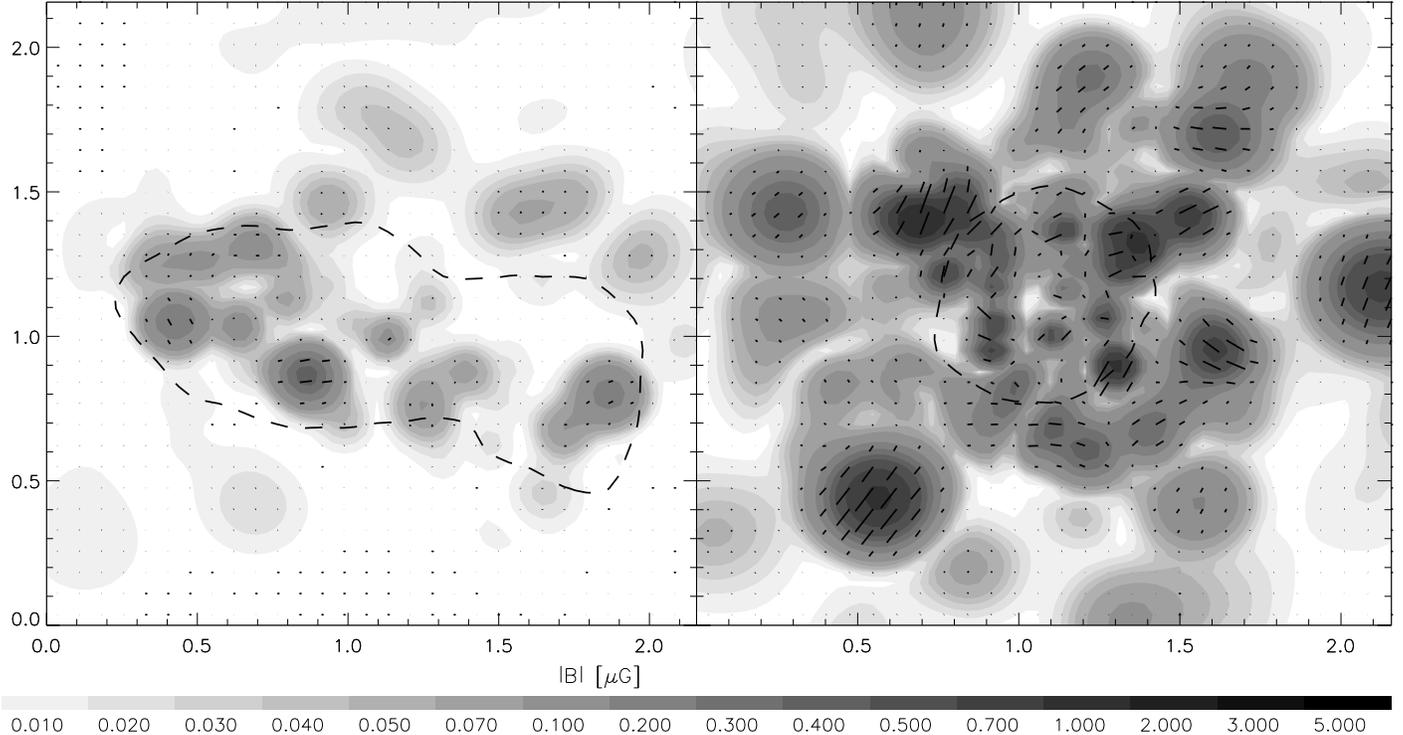}}
\caption{Slices cut through the centres of two simulated clusters are
  shown here. The gray-scale encodes absolute field strength $|\V{B}|$
  as indicated below the panels. The arrows mark magnetic field
  vectors in the planes of the slices. Notice that lengths, field
  strengths, and magnetic field vectors are scaled in the same way in
  both panels. The contour encloses the region emitting 90\% of the
  slices' total X-ray luminosity. The cluster shown in the left panel
  is highly substructured, as can be seen from the elongated X-ray
  contour. It consists of three main clumps, merging along a common
  line. Therefore, the X-ray region is more extended than in the more
  compact cluster shown in the right panel, even though the cluster in
  the right panel has twice the mass. This more massive cluster has a
  stronger magnetic field, due to the larger amount of gas compression
  during its formation.}
\label{fig:sheet}
\end{figure*}

Figure~\ref{fig:sheet} shows slices cut through the centres of two of
our simulated clusters with different masses. The magnetic field
develops a fluctuating pattern of regions with varying field strength,
indicated by the gray-scale.

The fluctuation amplitude clearly increases towards the cluster
centres, while the pattern gradually disappears away from the
clusters, as is best seen in the left panel. The cluster shown in the
right panel exhibits the same behaviour outside the plotted
region. The projection of the magnetic field vectors into the slices
are also shown as arrows in this illustration. They indicate that the
field changes on scales substantially smaller than the cluster
scale. The contour encloses the region which emits 90\% of the X-ray
flux, hence it surrounds the central bodies of the clusters in the
panels. The axes are scaled in Mpc.

\subsection{Radio and Inverse-Compton Emission}

The resolution in our present simulations is not good enough to
accurately describe the radio halo of a simulated cluster. Our
simulations reliably describe the mean field in clusters only on the
resolution scale and above, but we miss any smaller-scale
perturbations. For the radio emission, such small-scale fluctuations
are important. While small-scale field variations tend to cancel out
in the rotation measure because of the line-of-sight integration
involved, no such cancellation occurs with the radio
emission. Therefore, the mean magnetic field produced by our
simulations only provides a lower limit to the radio emission. We will
investigate this point in more detail as soon as higher-resolution
simulations will become available after an upgrade of our
combined Grape-workstation hardware.

So far, we have only performed a consistency check. We assumed that
the relativistic electron density is a constant fraction of the
thermal electron density, and that the electrons have a power-law
spectrum with exponent chosen such as to match the typical exponent of
observed radio spectra. The factor between the relativistic and
thermal electron densities was chosen such as to reproduce typical
radio luminosities; we find it to be of order $10^{-4}$. The question
then arises whether such a relativistic electron population would
produce unacceptable hard X-ray luminosities through their
inverse-Compton scattering of CMB photons. We adapted parameters such
as to meet measurements obtained in the Coma cluster (Kim et
al.~1990).

We calculated the inverse-Compton emission to find it negligible
compared to thermal emission in the soft X-ray bands, which in our
example agrees very well with the Coma measurements (Bazzano et
al.~1990). As mentioned before, a detailed investigation can only be
done with higher-resolution simulations. Nevertheless, this adds to
the consistent picture of the clusters in our simulation.

\subsection{Faraday Rotation}

One way to infer the mean magnetic field strengths in clusters is
provided by the Faraday rotation measure. Suppose the magnetic field
is randomly oriented and coherent within cells of size $l$. Let
$\bar{B}$ be the mean field strength, and $\bar{n}_{\rm e}$ the mean
electron density in a cell. The number of cells along the
line-of-sight is $\sim l/L$, where $L$ is the extension of the
cluster. Then, the expected {\em rms\/} rotation measure is
\begin{equation}
  \langle{\rm RM}^2\rangle^{1/2} \approx \frac{1}{3}\,812\,
  \frac{{\rm rad}}{{\rm m}^2}\,\left(\frac{l}{L}\right)^{1/2}\,
  \frac{\bar{n}_{\rm e}}{{\rm cm}^{-3}}\,
  \frac{\bar{B}}{\mu{\rm G}}\,
  \frac{l}{{\rm kpc}}\;.
\end{equation}
This estimate can be refined by assuming a density profile instead of
a constant mean electron density (Taylor \& Perley 1993). As long as
our simulations properly resolve the coherence scale $l$, the same
argument applies to rotation measures obtained from cluster
simulations. Hence, observed and synthetic Faraday-rotation
measurements provide a sensible way to compare the simulated to
observed clusters.

\begin{figure}[ht]
  \center{\includegraphics[width=\hsize]{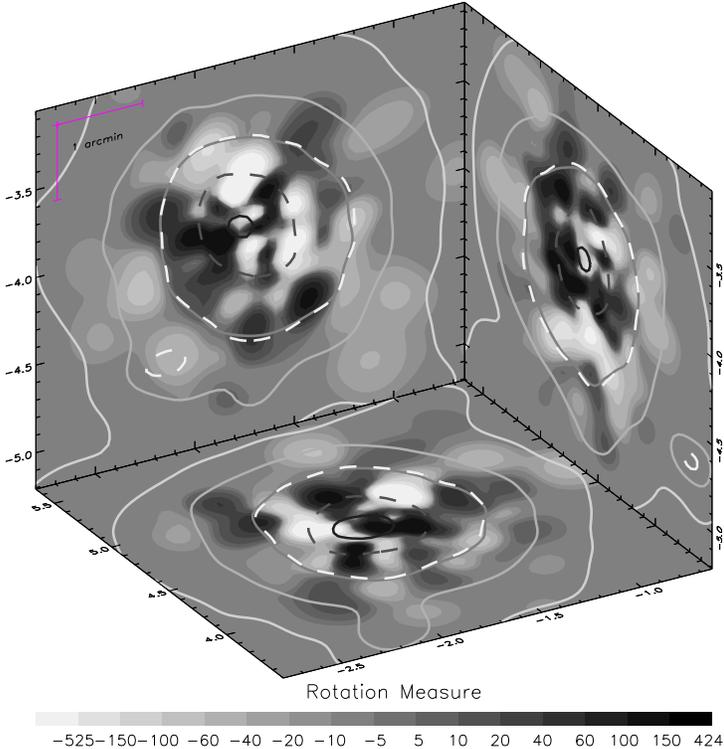}}
\caption{One of the most massive simulated clusters is shown in a
  three-dimensional box. The Faraday rotation measures produced by the
  cluster in the three independent spatial directions are projected
  onto the box sides and encoded by the gray-scale as indicated below
  the box. The gray solid curves are projected density contours as
  mentioned in the plot, and the gray dashed curve follows half the
  central density. The white dashed curve encompasses the region
  emitting 90\% of the projected X-ray luminosity. The cluster is
  taken from output data at redshift $z=0.5$. The plotted coordinates
  are physical coordinates, the virial mass is
  $\approx20\times10^{14}\,h^{-1}\,M_\odot$, and the virial radius
  lies outside the box shown.}
\label{fig:3dbox}
\end{figure}

Among other things, Fig.~\ref{fig:3dbox} shows the rotation measure
for one cluster projected on the sides of its simulation cube. The
value of the rotation measure is encoded by the gray-scale, as
indicated in the plot. The white dashed line is the projected 90\% X-ray
contour, thus marking the central body of the cluster in the box.
Qualitatively, the simulated RM patterns look very similar to
observations as those in Coma shown in Fig.~\ref{fig:coma}. All
measurements of a significant RM for Coma fall within the Abell radius
(dotted circle). In observations and our simulations alike, they do
not much extend beyond the X-ray boundary. In both simulations and
observations, we find elongated regions in which the sign of the RM
does not change. We also find regions with vanishing rotation measure
next to regions with high rotation measure. Close to cluster centres,
the rotation measure typically changes sign on much smaller scales
than farther out. Coherent RM patches thus become smaller, and they
are separated by boundaries of negligible RM. The heights of RM peaks
decrease away from the cluster centre. This behaviour is illustrated in
the lower right panel of Fig.~\ref{fig:pltcum} for the Coma cluster
(points).

Patterns in the rotation measure can now statistically be compared
across different initial conditions in the simulations, or with
comparable statistics of observations.

\section{Statistics of Faraday Rotation Measures}

\subsection{Different Initial Field Set-Ups}

To evaluate what the two different kinds of initial magnetic-field
set-ups imply for the observations of rotation measures, we compute
rotation-measure maps from the cluster simulations and compare them
statistically. We use two methods for that. First is the usual
Kolmogorov-Smirnov test, which evaluates the probability with which
two sets of data can have been drawn from the same parent
distribution. Second is an excursion-set approach, in which we compute
the fraction ${\cal F}$ of the total cluster surface covered by RM
values exceeding a certain threshold. For the cluster surface area, we
take the area of the region emitting 90\% of the X-ray luminosity.

\begin{figure}[ht]
  \center{\includegraphics[width=\hsize]{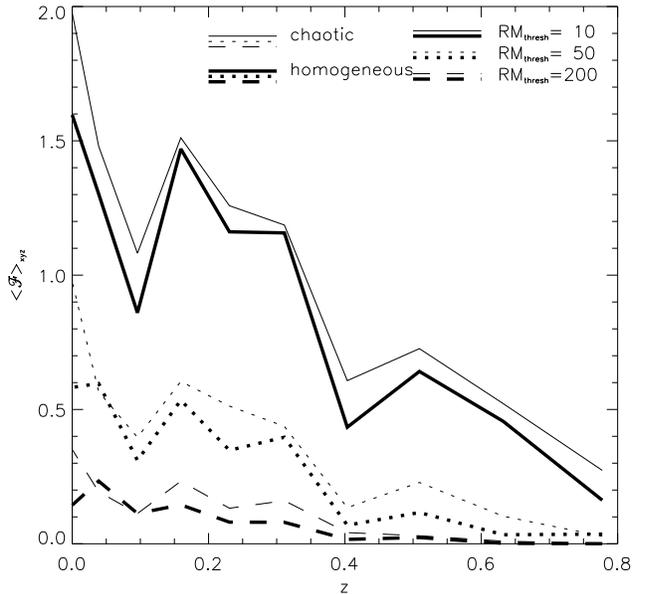}}
\caption{This figure shows the fraction ${\cal F}$ of the cluster area
  which is covered by regions in which the absolute value of the
  rotation measure exceeds a certain threshold. The cluster area is
  taken to be the area of the region emitting 90\% of the X-ray
  luminosity. The results are averaged over the three independent
  spatial directions. Line types distinguish different RM thresholds,
  as indicated in the figure.}
\label{fig:bild}
\end{figure}

Figure~\ref{fig:bild} shows how the RM distributions evolve in
clusters in which the initial magnetic field was set up either
homogeneously or chaotically. We use the excursion-set approach here,
i.e.~we plot the fraction ${\cal F}$ of the cluster area that is
covered by regions in which the RM exceeds a certain threshold.

Quite generally, this fractional area for fixed RM threshold increases
during the process of cluster formation. There is only little
difference between initially homogeneous (heavy lines) and chaotic
(thin lines) magnetic fields. At later stages of cluster evolution,
the area covered by a minimum rotation measure tends to decrease
again. This general behaviour is an effect of sub-structures in
clusters: While the clusters form, infalling material creates shocks
in the intra-cluster gas, where the magnetic field is amplified by
compression, and the motion of accreted sub-lumps through the
intra-cluster plasma builds up shear flows which stretch the magnetic
field. As the clusters relax, the magnetic field strengths decrease
somewhat because of dissipation.

Another way of analysing these patterns is to randomly shoot beams of
a certain width through the clusters, determine the average rotation
measure in these beams and statistically compare such synthetic
Faraday-rotation measurements to observed samples.

\begin{figure}[ht]
  \center{\includegraphics[width=\hsize]{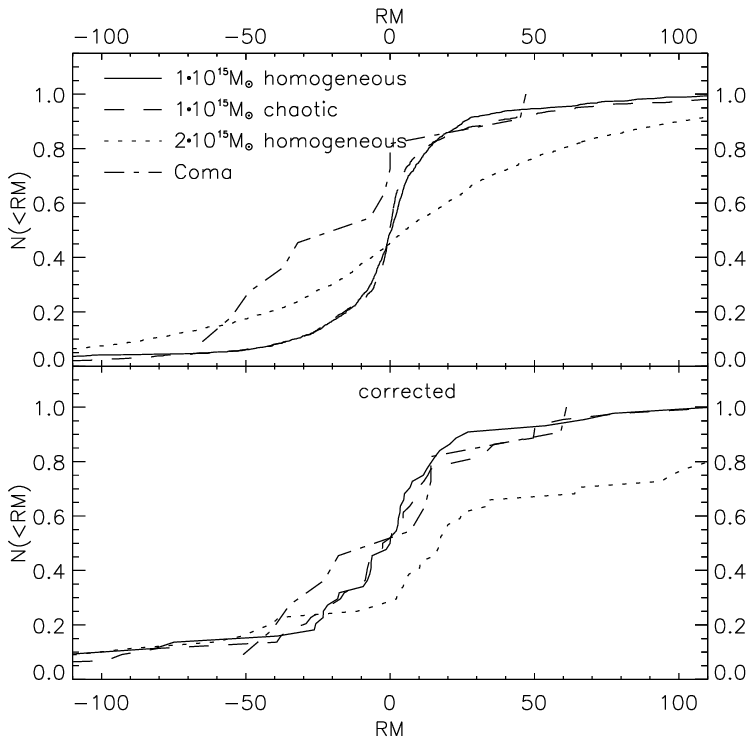}}
\caption{Four cumulative rotation-measure distributions are shown in
  the upper panel. Solid and dashed curves: A cluster model with final
  virial mass $M_{200}=10^{15}\,M_\odot$ was simulated twice, once
  with a homogeneous (solid curve) and once with a chaotic (dashed
  curve) initial magnetic field of the same {\em rms\/} strength. At
  redshift zero, synthetic rotation-measure samples were obtained from
  both clusters by determining the Faraday rotation along $10^4$ beams
  randomly shot through the clusters. Evidently, the two curves are
  very similar, indicating that the vastly different magnetic-field
  set-ups do not lead to any significant difference in the observable
  RM distributions for a given cluster. For comparison, the dotted
  curve shows similar results for a cluster with twice the mass, in
  which the initial magnetic field was homogeneous. The RM variance is
  substantially increased. This indicates that RM distributions are
  expected to vary much more strongly with cluster mass than with the
  initial field configuration. Finally, the dash-dotted curve shows
  the distribution of the measurements obtained in the Coma
  cluster. Various corrections detailed in Sect.~\ref{sec:coma} are
  applied to the latter distribution in the lower
  panel. Kolmogorov-Smirnov tests show that the Coma measurements are
  fully compatible with the synthetic RM distributions in the less
  massive cluster, and incompatible at the 99\% level with the
  simulations in the more massive cluster.}
\label{fig:pltcum}
\end{figure}

We first pick a cluster of $10^{15}\,M_\odot$ final virial mass,
simulated with both homogeneous and chaotic initial magnetic field
configurations, and randomly shoot $10^4$ beams through it, along
which we compute the Faraday rotation measure. The resulting
cumulative RM distributions are shown in Fig.~\ref{fig:pltcum}, where
the solid and dashed curves show results for the homogeneous and
chaotic magnetic-field set-up, respectively. Evidently, there is very
little difference between these two distributions, indicating that the
vastly different initial magnetic-field configurations lead to
virtually indistinguishable rotation-measure samples.

To emphasise this statement, we add to Fig.~\ref{fig:pltcum} a
short-dashed curve showing the RM distribution obtained from a cluster
with twice the final mass, in which the initial magnetic field was set
up homogeneously. The curve is much flatter than for the lower-mass
cluster, showing that the {\em rms\/} rotation measure is
substantially larger in the more massive cluster. Obviously, clusters
with different masses produce much more different RM distributions for
the same class of magnetic initial conditions than clusters of the
same mass do for different classes of initial conditions. Hence, the
scatter of results across a sample of clusters with different masses
is expected to be much too large to allow to distinguish in principle
between the two kinds of magnetic initial conditions.

As an additional difficulty, the number of sources with sufficiently
high, sufficiently polarised flux behind a given cluster is fairly
limited. Hence, the small number of possible beams measuring RM values
in clusters further restricts the power of statistical comparisons
between cluster magnetic fields.

\subsection{Varying Initial Field Strengths}

We now investigate whether this result holds true for a sample of
simulated clusters, irrespective of their masses and also their
initial magnetic field strengths. For that purpose, we produce a
sample of 13 clusters, each of which is simulated twice, once with
homogeneous and once with chaotic initial magnetic field. To increase
the scatter in mass, we take the clusters at output redshifts between
$z=0.78$ and $z=0$. The cluster masses then range within
$(8-25)\times10^{14}\,M_\odot$. Initial {\em rms\/} magnetic field
strengths are varied between $(1/3-3)\times10^{-9}\,{\rm G}$.

As before, we then compute RM distributions from $10^4$ rays shot
randomly through the clusters inside the X-ray boundary, and compare
them with a Kolmogorov-Smirnov test. The results show that even 30 RM
values per cluster are insufficient to distinguish between the two
field set-ups at the 3-$\sigma$ level for all simulated clusters. In
particular, in the Coma cluster, which is the best-observed cluster in
that respect, there are only ten RM values available within the X-ray
boundary.

An important result of these comparisons is that there is no
statistically significant difference between the RM patterns exhibited
by clusters whose initial magnetic fields were either homogeneous or
chaotic. This illustrates that the final field configuration in the
cluster is dominated by the cluster collapse rather than the initial
field configuration. Approximately on the free-fall time-scale, the
initial field is first ordered more or less radially, and when the
cluster relaxes later, the field is stirred by shear flows in the
intra-cluster gas. Although we started two different sets of
simulations with extreme, completely different initial conditions, the
resulting fields cannot significantly be distinguished by observations
of Faraday rotation. We believe that this scenario is typical for
hierarchical models of structure formation, but may be different in
e.g.~cosmological defect models.

\section{Comparison with observed data}

\subsection{\label{sec:coma} Individual clusters}

\begin{figure}[ht]
  \center{\includegraphics[width=\hsize]{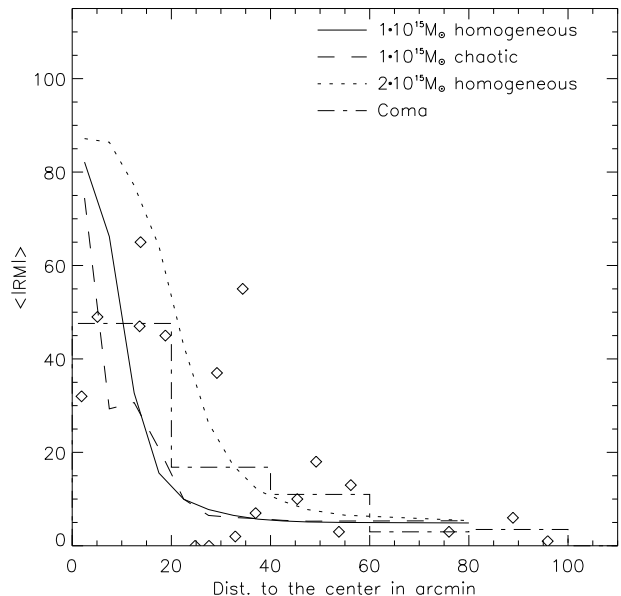}}
\caption{Several radial distributions of simulated and observed
  rotation measures are shown. RM values observed in Coma are
  represented by diamonds and the dash-dotted histogram. The solid and
  long-dashed distributions were obtained from a simulated cluster
  with mass $\approx10^{15}\,M_\odot$, for homogeneous and chaotic
  initial magnetic fields, respectively. For the dotted curve, a
  cluster with approximately twice the mass was used.}
\label{fig:pltrad}
\end{figure}

We can now compare the simulated RM statistics with measurements in
individual clusters like Coma or Abell~2319, for which a fair number
of measurements is available. The distribution of the RM values as a
function of distance to the cluster centre are shown for the Coma
cluster as points in Fig.~\ref{fig:pltrad}. These measurements have
the cumulative distribution (dash-dotted curve) shown in
Fig.~\ref{fig:pltcum}. This distribution can now be compared to the RM
values obtained from the simulations (other lines in
Fig.~\ref{fig:pltrad}). For a fair comparison, we have to take the
radial source distribution on the sky relative to the cluster centre
into account. This can easily be achieved by choosing random beam
positions under the constraint that their radial distribution match
the observed one. In our example, the Kolmogorov-Smirnov likelihood
for 30 simulated beams to match the ten measurements in Coma falls
between 30\% and 50\% for the less massive cluster (solid and dashed
curves), compared to a few per cent for the more massive one (dotted
curve). Note that the less massive cluster has approximately the
estimated mass of Coma.

Figure~\ref{fig:pltcum} indicates that the cumulative RM distribution
in Coma is not symmetric about the origin ${\rm RM}=0$, implying
$\langle{\rm RM}\rangle\ne0$ in Coma. It is possible to apply a
constant additive correction of $\approx14\,{\rm rad\,m}^{-2}$ to the
Faraday-rotation measurements in Coma to achieve a vanishing mean
$\langle{\rm RM}\rangle=0$. This additive correction yields the
dash-dotted curve in the lower panel of Fig.~\ref{fig:pltcum}, labelled
as ``corrected''.

In addition, we can statistically correct for Faraday rotation
intrinsic to the sources using measurements near the Coma cluster, but
outside the X-ray boundary. For the simulated RM distributions in the
upper panel of Fig.~\ref{fig:pltcum}, this procedure yields the curves
in the lower panel, which can now directly be compared with the
dash-dotted line from the corrected Coma measurements. The
Kolmogorov-Smirnov likelihood for simulations and observations to
agree is then increased by a factor of $\approx2$ for the particular
clusters used.  Nonetheless, we only apply those corrections where
explicitly mentioned. If we use the estimated masses of Coma or A~2319
to select simulated clusters, we found that for an initial field of
$\approx10^{-9}\,{\rm G}$ at $z=15$ we achieve excellent statistical
agreement with the simulated RM distributions.

A more detailed comparison using a sample of simulated clusters with
different initial magnetic-field energies leads to the intuitive
result that simulated clusters with smaller initial magnetic field
need to have larger masses to achieve reasonable agreement with the
Coma measurements.

An alternative way to compare the measurements to the simulations is
to radially bin the absolute value of RM around the cluster
centre. One can then calculate the mean absolute value of RM, the
dispersion, or other characteristic quantities of the RM distribution
in dependence of radius.

This is illustrated in Fig.~\ref{fig:pltrad}, where the mean absolute
RM value is plotted for the same three clusters as in the previous
figures. As mentioned before, due to the intrinsic RM values of the
sources, there is a signal left in the observations for $\phi>40'$.
To match the observations at this angular scale, we apply the additive
correction of $7.7\,{\rm rad\,m}^{-2}$, which is the mean absolute RM
value for the observed control sample.  This value is added to all
theoretical curves. The Faraday rotations measured in the Coma cluster
(circles) are inserted according to their distance to the cluster
centre. Additionally, the radial distribution of the measurements is
shown by a histogram (dash-dotted line). Clearly, for the less massive
cluster (solid and dashed curves), the agreement with the observations
is better than for the more massive one (dotted line), for which the
RM reaches too high values near the centre. It is also clearly visible
that the difference between the homogeneous and chaotic initial
conditions for the magnetic field is insignificant.

\subsection{A sample of clusters}

\begin{figure}[ht]
  \center{\includegraphics[width=\hsize]{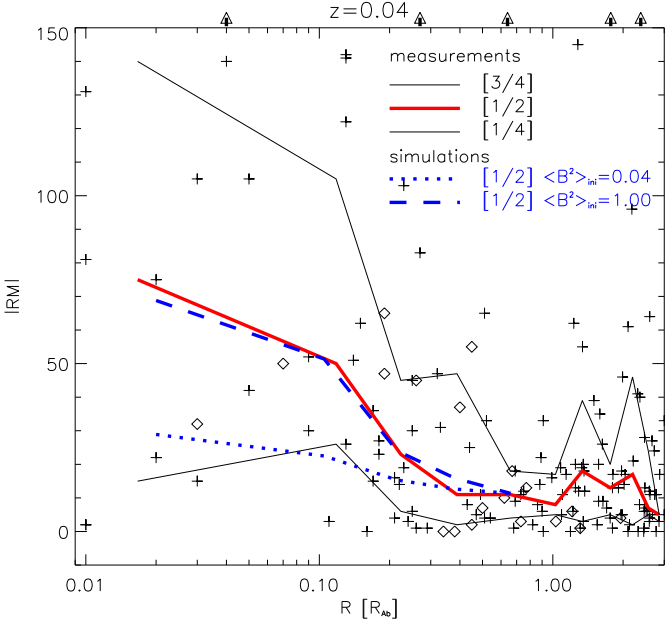}}
\caption{Comparison of simulated results with a sample of measurements
  in Abell clusters. The absolute values of Faraday rotation
  measurements (Kim et al.~1991) vs.~radius in units of the Abell
  radius are shown. Obviously, the dispersion increases towards the
  cluster centre. The solid curves mark the median and the 25- and
  75-percentiles of the measurements, and the dashed and dotted curves
  are the medians obtained from simulated cluster samples starting
  with high and low magnetic fields, respectively. Specifically,
  $\langle B^2\rangle^{1/2}=10^{-9}\,{\rm G}$ and
  $0.2\times10^{-9}\,{\rm G}$ were chosen for the initial {\em rms\/}
  magnetic field strength. The radial bins are chosen such as to
  contain 15 data points each. The scatter in the observations is
  large, but the simulations with the stronger initial magnetic field
  seem to better match the observations.}
\label{fig:8}
\end{figure}

Unfortunately, in most clusters the number of radio sources suitable
for measurements of Faraday rotation is only one or two. We therefore
also compare a sample of simulated clusters with a compilation of RM
observations in different clusters. Figure~\ref{fig:8} shows the
values published by Kim et al.~(1991). We take the absolute RM value
and plot it against the distance to the nearest Abell cluster in units
of the estimated Abell radius (crosses). Measurements in the Coma
cluster are marked with diamonds.

The increase of the signal towards the cluster centres is evident. In
order not to be dominated by a few outliers, which in some cases even
fall outside the plotted region and are marked by the arrows at the
top of the plot, we did not calculate mean and variance, but rather
the median and the $25$- and $75$-percentiles of the RM distribution
(solid curves). We chose distance bins such that each bin contains the
same number (15) of Faraday-rotation measurements. We checked that
reducing this number to 10 did not change the curves at all.

The median of synthetic radial rotation-measure distributions obtained
from our simulations for different initial magnetic field strengths is
shown by the dashed and dotted curves. The initial magnetic field
strengths for the simulation are indicated by the different line types
as indicated in the figure.

In order to correct our synthetic results for source-intrinsic Faraday
rotation, we calculate the mean signal in the distance range between
one and three Abell radii, and add the resulting $\approx16\,{\rm
rad\,m}^{-2}$ to the simulated curves. Figure~\ref{fig:8} shows that
an initial field strength of order $\approx10^{-9}\,{\rm G}$ can
reasonably reproduce the observations. We took the simulation at a
redshift of $z=0.04$ for this plot, but the results do not change
substantially if we take the output at $z=0.5$ to mimic a differently
normalised cosmological model.

\section{Conclusions}

We performed cosmological simulations of galaxy clusters containing a
magnetised gas. At redshift $15$, initial magnetic fields of
$\approx10^{-9}\,{\rm G}$ strength were set up, and their subsequent
evolution was calculated solving the hydrodynamic and the induction
equations assuming perfect conductivity of the intra-cluster
plasma. The special-purpose hardware ``Grape'' was used to calculate
the gravitational forces between particles, and the hydrodynamics was
computed in the SPH approximation. A previously existing code
(GrapeSPH; Steinmetz 1996) was employed and modified to incorporate
the induction equation and magnetic forces.

Extensive tests of the code show that it well reproduces test problems
and that the divergence of the magnetic field remains negligible
during all phases of cluster evolution.

The primary purpose of the simulations was to address two
questions. (i) Given primordial magnetic fields of speculative origin
and strengths of $\approx10^{-9}\,{\rm G}$, can simulations reproduce
Faraday-rotation observations in galaxy clusters, from which field
strengths of order a $\mu{\rm G}$ are inferred? (ii) Is the structure
of the primordial fields important, i.e.~does it matter for the
results whether the fields are ordered on large scales or not? We can
summarise our results as follows:

\begin{enumerate}

\item Almost everywhere in clusters, the magnetic fields remain
  dynamically unimportant. Generally, the Alfv\'en speed is at most
  $\approx10$ per cent of the sound speed.

\item While simple spherical collapse models predict final cluster
  fields of $\approx10^{-7}\,{\rm G}$, fields of $\mu{\rm G}$ strength
  are achieved in the simulations. This additional field amplification
  is due to shear flows in the intra-cluster plasma. Detailed
  investigations show that particularly strong fields arise in eddies
  straddling regions of gas flows in different directions.

\item Cumulative statistics of synthetic Faraday-rotation measurements
  reveal that the cluster simulations reproduce observed RM
  distributions very well. Generally, more massive clusters require
  somewhat weaker initial magnetic fields to reproduce observed RM
  values, but initial field strengths of order $10^{-9}\,{\rm G}$
  yield results excellently compatible with observations.

\item Two extreme cases for the structure of the initial fields,
  namely either completely homogeneous or completely chaotic field
  structures, lead to final field configurations which are entirely
  indistinguishable with Faraday-rotation measurements. It is
  therefore unimportant for interpreting RM observations in clusters
  {\em how\/} primordial fields are organised as long as they have the
  right mean strength. The field evolution during cluster formation is
  thus much more important for the final field structure than any
  differences in the primordial field structure.

\end{enumerate}

The last item has an important consequence for the epoch when
primordial fields need to be in place. Cluster formation is a
cosmologically recent process, i.e.~it sets in at redshifts below
unity. For the field structure in clusters, it is therefore only
important that sufficiently strong seed fields be present when cluster
formation finally sets in at cosmologically late times. The earlier
history of the seed fields is not accessible through cluster
observations because information on the primordial field structure is
erased during cluster collapse.

We note that our simulations emphasise the role of systematic
large-scale flows, i.e.~shear flows which are generic constituents of
gravitationally unstable multifluid-systems, like collapsing galaxy
clusters. Any plasma motion with sufficient velocity gradient will
stretch, entangle and disentangle the magnetic field, which is
amplified thereby, and since the shear flows in clusters are
three-dimensional, the field amplification is a global
effect. However, even these intrinsic shear flow patterns have not
amplified the field strength to a value where it would be dynamically
important; the field is not in equipartition at any stage of the
cluster evolution.

An important weakness of our simulations so far is the limited spatial
resolution. While we believe that the resolution is sufficient for
producing synthetic Faraday-rotation measurements because there
magnetic fields are integrated along the line-of-sight so that
small-scale fluctuations are averaged out, simulations of the radio
emission due to relativistic electrons in the cluster field require
higher resolution. We can therefore not use our present simulations to
reliably reproduce cluster radio haloes. Nevertheless, we verified
using simple assumptions that our cluster simulations meet
observational constraints on the hard X-ray emission due to CMB
photons which are Compton-upscattered by relativistic electrons. We
also note that the field amplification due to shear flows may be
somewhat underestimated due to the limited resolution.

The resolution of our simulations will be increased in the near
future. They can (and will) then be applied to a number of additional
questions related to magnetic fields in clusters, like the structure
of radio haloes, the importance of magnetic reconnection for the heat
balance in cluster cores, the structure of magnetised cooling flows
and so forth.

Finally, we would like to discuss the general lack of diffuse cluster
radio haloes (e.g. Burns et al. 1992). We now know that most galaxy
clusters contain intracluster fields which are comparable with
galactic magnetic field strengths of several $\mu$G (e.g.~Kronberg
1994 and references therein), but only a few of them reveal a diffuse
radio halo. Why? In view of our simulations this problem is even more
severe, since according to our results all galaxy clusters will
produce fields of $\mu$G strength, simply by collapse and the
accompanying shear flows. Thus, every cluster should shine in the
radio.

Here, we would like to present a possible explanation for the missing
cluster radio haloes in terms of inefficient particle acceleration in
the cluster medium. Our hypothesis is: there is in general no extended
radio emission in haloes of galaxy clusters since the number of
relativistic electrons is too small, i.e.~the particles are not
significantly re-accelerated in the cluster.

We can support this hypothesis with two basic arguments:

First, the galaxies in galaxy clusters experience lots of tidal
interactions, which excite star formation (SF). Such SF-episodes heat
up galaxy haloes and inflate them via galactic winds, which contain
cosmic rays and strong magnetic fields (an example par excellence is
M82). Thus, these galaxies will slow down the relativistic electrons
by considerable synchrotron losses; the energy-loss time $t_{\rm
loss}\propto\gamma^{-1}B^{-2}$ decreases with increasing field
strength. Therefore, whatever relativistic electron population reaches
intergalactic space has already lost a substantial fraction of its
energy. Furthermore, the strong shear flows in the intracluster plasma
close to the galaxies amplify the halo fields, increase the
synchrotron losses and concentrate the high energy particles to the
close neighbourhood of the galaxies themselves.

Second, the only possibility to create an extended radio halo would be
electron re-acceleration by shock waves or turbulence (Schlickeiser et
al.~1987). But, as investigated by Dorfi and V\"olk (1996), the
efficiency of Fermi acceleration of supernova-driven shock waves in hot
and tenuous plasmas is drastically reduced. They studied the problem
why elliptical galaxies, while containing strong magnetic fields
(known from RM-measurements) and enough supernova shocks, do not
exhibit significant radio emission. The main reason for the decrease
in shock wave acceleration in hot plasmas is the decrease of the Mach
number and the decrease of the shock power before the Sedov phase.

Translating their results into the context of galaxy clusters, we
would draw the following picture: Relativistic electrons originate in
galaxies, they hardly overcome the magnetic barriers of the galactic
halo fields. Due to the high sound speed, the shock waves in the
cluster medium have only small Mach numbers, and they do not
significantly accelerate. The efficiency of MHD turbulence for particle
acceleration is also lower in hot, tenuous plasmas since the
fluctuations are damped by the hot ions (Holman et al.~1979).

Only external influences like further accretion, especially of cold
material, is able to raise the Mach numbers to get significant
particle re-acceleration in the clusters (see also Tribble 1993).

\section*{Acknowledgements}

We wish to thank the referee, A.~Shukurov, for his valuable and
detailed comments which helped to clarify and improve the paper
substantially.

\end{document}